# COMMENT ON "PHASE FLUCTUATION SUPERCONDUCTIVITY IN OVERDOPED La$_{2-x}$Sr$_x$CuO$_4$".

Jesús Mosqueira, Manuel V. Ramallo, Félix Vidal

LBTS, Facultade de Física, Universidade de Santiago de Compostela,

ES-15782 Spain

By measuring the in-plane resistivity as a function of temperature and magnetic field, $\rho_{ab}(T,H)$, in overdoped La$_{2-x}$Sr$_x$CuO$_4$ (LSCO) superconductors, in a recent Letter Rourke and coworkers[1] have addressed the long standing but still open issue of the interplay between pseudogap and superconducting fluctuations in cuprates. Although in their Letter Rourke et al. recognize that their samples could be affected by "*strong disorder inherent in LSCO*", unfortunately these authors have analyzed their interesting measurements without taking into account the corresponding intrinsic-like inhomogeneities of the zero-field critical temperature $T_{c0}$. In fact, the unavoidable presence in non-stoichiometric cuprates of $T_c$-inhomogeneities associated with the random distribution of dopants, and their dramatic influence on the measurements of the electrical resistivity and the magnetization around their superconducting transition, was already stressed in different works[2,3]. In addition, it was shown that in LSCO these intrinsic-like $T_c$-inhomogeneities have characteristics lengths much larger than the in-plane superconducting coherence length amplitude[4]. This is a central result because, then, the own nature of the superconducting fluctuations above $T_c$ is not affected by the $T_c$-inhomogeneities, these last affecting the measurements *after* the intrinsic fluctuation effects[2-4]. By just focusing on an example, here we will indicate how these $T_c$-inhomogeneities may be easily taken into account in the data of Ref.1. Our results strongly suggest then conclusions just opposite to those stressed in that paper, but similar to those earlier proposed by Currás et al. in Ref. 2 and later confirmed by different authors[5,6]: the adequacy of the conventional Gaussian-Ginzburg-Landau (GGL) approach to describe at a phenomenological level the paraconductivity in cuprates, independently of doping, instead of the popular phase-fluctuation scenario (see also Ref. 7 and our Comment on that paper).

One may illustrate the influence of these $T_c$-inhomogeneities with long characteristic lengths, that will be present even in ideal overdoped LSCO crystals[4], through the onset temperature, $T_2(p)$, for the superconducting fluctuations as a function of the nominal hole doping, p. The $T_2(p)$ data of Fig. 2d of Ref. 1 in the overdoped regime are reproduced here in Fig. 1 (circles), together with those earlier measured by Currás et al.[2] in LSCO films (triangles), and presented in Table I and in Fig. 8b of Ref. 2. The independence of $T_2/T_{c0}$ of the doping level shown by the data of Ref. 2 was confirmed in other cuprate superconductors by other authors by using different procedures to estimate $T_2$, including the quenching of superconductivity by high magnetic fields (see, e.g., Refs. 5-7 and references therein). Therefore, this central result may be considered as "model independent". The solid line in Fig.1 corresponds to $T_2(p) = 1.7 T_{c0}(p)$, as predicted by the Gaussian-Ginzburg-Landau approach for superconducting fluctuations, extended to high reduced temperatures by introducing the so-called total-energy cutoff (extended GGL approach).[2,8] As such GGL onset temperature corresponds to homogeneous superconductors, it is labeled $T_2^{hom}$ in Fig. 1.

The origin of the discrepancies observed in Fig. 1 may be attributed to the presence of $T_c$-inhomogeneities associated with chemical disorder, which will deeply affect the temperature location of the $d\rho_{ab}/dT$ minimum above $T_{c0}(p)$, the procedure used in Ref. 1 to determine $T_2(p)$ (see Figs. 1d and S2-S4 in Ref.1). Without an adequate criteria, in presence of $T_c$-inhomogeneities this procedure may strongly overestimate $T_2$ (see, e.g., Ref. 5, where these spurious effects were mitigated by analyzing the behavior of $d^2\rho_{ab}/dT^2$). Note that near optimum doping the $d\rho_{ab}/dT$ minimum will be less affected by chemical disorder, due to the lower $dT_{c0}/dp$, explaining the agreement shown in Fig. 1 at those doping levels.

The simplest way to illustrate how even the intrinsic-like chemical disorder may deeply affect the $T_2/T_{c0}$ data of Rourke et al. is by combining the corresponding GGL relationship for homogenous superconductors[2,8] with the results of Ref. 4 for the intrinsic dispersion in the p value, $\Delta p$ (FWHM), associated with the random distribution of dopants in LSCO [Eq. (5) of Ref. 4]. For that, we have taken into account that the volume fraction of domains with hole dopings above $p+2\Delta p$ or below $p-2\Delta p$ in the Gaussian p-distribution represents only one part in $10^6$, and their effects remain below the resolution of the resistivity measurements. This leads to the dashed curve in Fig.1, whereas the dot-dashed curve is just $1.7T_{c0}(max)/T_{c0}(p)$, where $T_{c0}(max) \approx 38$ K is the maximum $T_{c0}$ in the LSCO system. As shown in Fig. 1, the $T_2/T_{c0}$ data-points of Ref. 1 are well between these two extreme curves, denoted as $T_2^{min}/T_{c0}$ and, respectively, $T_2^{max}/T_{c0}$. This last result is consistent with the fact that the resistive transition widths that may be inferred from Figs. 1c and S2-S4 of Ref. 1 are larger than the intrinsic ones. Note that a similar result could be obtained independently of the GGL approach, by just assuming a linear relationship between $T_2$ and $T_{c0}$ in absence of any chemical disorder.

In conclusion, the straightforward analysis we have presented here shows that if the effects associated with chemical disorder are taken into account, the measurements of Rourke et al.[1] of the onset temperature for the superconducting fluctuations in the overdoped regime will lead to conclusions just opposite to those stressed in that paper but, as already noted in the introduction of this Comment, similar to those earlier proposed by Currás et al. in Ref. 2 and later confirmed by different authors:[5-7] the independence of $T_2/T_{c0}$ of the hole doping, well within the experimental uncertainties, a result that does not support the popular phase-fluctuation scenario. Complementarily, our present analysis also shows that the *actual* $T_2$ that one may infer from the measurements of Rourke et al. may be accounted for by the presence of GGL-type fluctuations up to about twice the local values of $T_{c0}$. Also, the extended GGL scenario leads to a $T_2(H)$ dependence (see, e.g., Ref. 9) compatible with the one obtained in the remaining measurements of Rourke et al.[1] under high magnetic fields. Note finally that the measurements of any observable around $T_c$ in any non-stoichiometric cuprate superconductor (included, obviously, in the overdoped Y-based and Tl-based crystals, also briefly commented in Ref. 1) will also be affected by the intrinsic-like chemical disorder associated with the random distribution of the dopant ions.[2-4,10] So, the observation in these overdoped compounds of a $T_2(p)$ behavior similar to the one reported for LSCO in Ref. 1 will just confirm the presence of these unavoidable $T_c$-inhomogeneities.

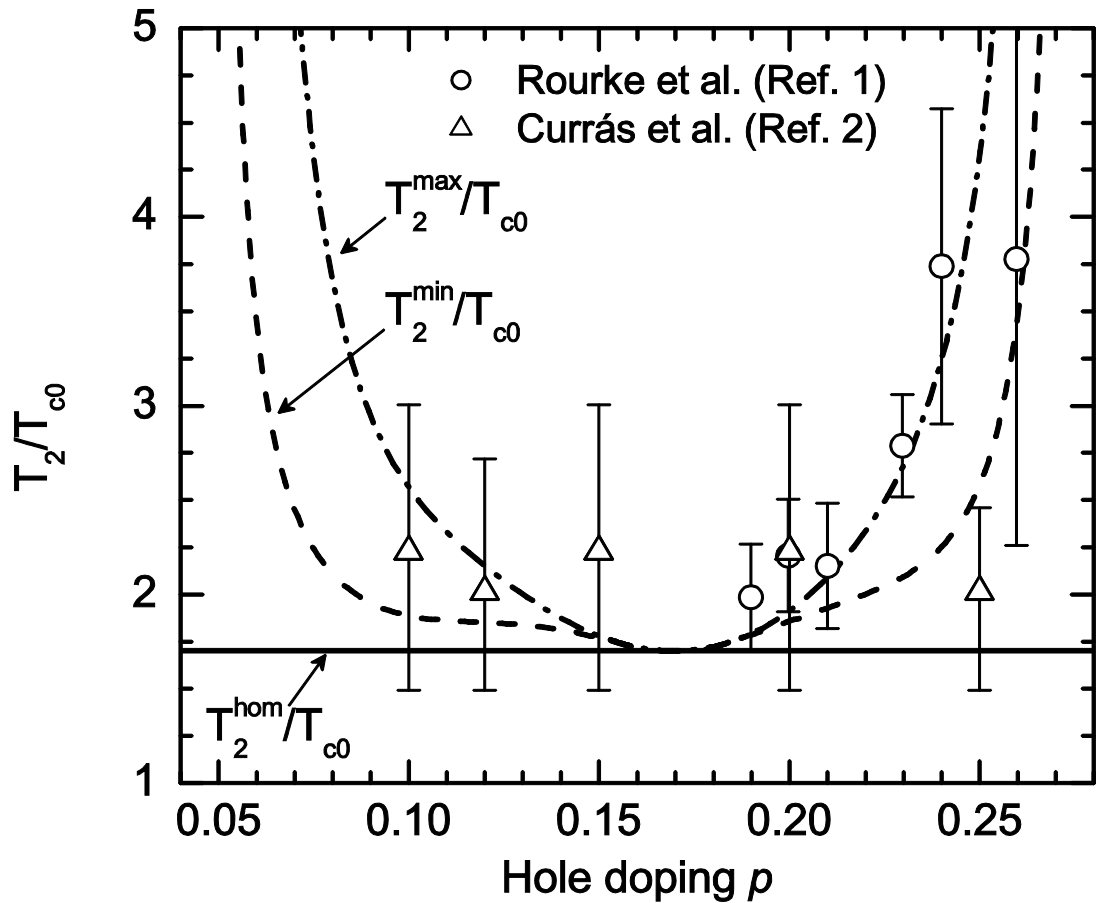

Fig. 1. Relative temperature onset for the superconducting fluctuations, as a function of hole doping, measured in overdoped LSCO crystals in Ref. 1 (circles). The triangles are the earlier results in LSCO films of Currás et al. (Ref. 2). The curves were obtained by taking into account both the superconducting fluctuations, through the extended GGL approach, and the chemical disorder effects on $T_c$.